\documentclass[conference]{IEEEtran}
\IEEEoverridecommandlockouts
\usepackage{cite}
\usepackage{amsmath,amssymb,amsfonts}
\usepackage{algorithmic}
\usepackage{graphicx}
\usepackage{textcomp}
\usepackage{xcolor}
\usepackage{todonotes}
\usepackage{subfig}
\usepackage{hyperref}
\newcommand\copyrighttext{%
 \footnotesize \textcopyright  2021 IEEE.  Personal use of this material is permitted.  Permission from IEEE must be obtained for all other uses, in any current or future media, including reprinting/republishing this material for advertising or promotional purposes, creating new collective works, for resale or redistribution to servers or lists, or reuse of any copyrighted component of this work in other works. 
}
\newcommand\copyrightnotice{%
\begin{tikzpicture}[remember picture,overlay]
\node[anchor=south,yshift=10pt] at (current page.south) {\fbox{\parbox{\dimexpr\textwidth-\fboxsep-\fboxrule\relax}{\copyrighttext}}};
\end{tikzpicture}%
}
\def\BibTeX{{\rm B\kern-.05em{\sc i\kern-.025em b}\kern-.08em
    T\kern-.1667em\lower.7ex\hbox{E}\kern-.125emX}}
\begin{document}
\bstctlcite{IEEE:conf}
\title{Requirement analysis for an artificial intelligence model for the diagnosis of the COVID-19 from chest X-ray data\\
}

\author{\IEEEauthorblockN{Tuomo Kalliokoski}% \orcidlink{0000-0001-5291-5577}}
\IEEEauthorblockA{\textit{Faculty of Information Technology} \\
\textit{University of Jyväskylä},\\
Jyväskylä, Finland, \\
ORCID: 0000-0001-5291-5577,\\
tuomo.kalliokoski@jyu.fi}
}

\maketitle
\copyrightnotice

\begin{abstract}
There are multiple papers published about different AI models for the COVID-19 diagnosis with promising results. Unfortunately according to the reviews many of the papers do not reach the level of sophistication needed for a clinically usable model. In this paper I go through multiple review papers, guidelines, and other relevant material in order to generate more comprehensive requirements for the future papers proposing a AI based diagnosis of the COVID-19 from chest X-ray data (CXR). Main findings are that a clinically usable AI needs to have an extremely good documentation, comprehensive statistical analysis of the possible biases and performance, and an explainability module. 
\end{abstract}


\section{Introduction}
Ever since the World Health Organization classified the COVID-19 as a Public Health Emergency of International Concern (PHEIC) \cite{WHO_PHEIC}, which is more commonly called as a pandemic \cite{WHO_pandemic}, the AI field has produced a multitude of papers related to diagnosing the COVID from various data. Reviews which focus on the clinical suitability of the models presented like references \cite{Hryniewska2021}, \cite{Roberts2021}, \cite{Wynants2020}, \cite{Burlacu2020}, \cite{Albahri2020}, and \cite{Alghamdi2021} have been critical about various aspects of the reviewed papers. Thus there is a need for a proper requirements analysis for the AI based diagnosis of the COVID-19 from CXR data so that these shortcomings can be remedied.

This requirement analysis includes general ethical considerations, general AI model building considerations, and clinical considerations for both the AI in general and in the radiology diagnosis. For these considerations I will use relevant publications as sources of the requirements. Many of the other sources are reviews of previous work in which I will concentrate on the criticism of the analyzed publications. 

In the next section I will go through the selected sources in order to find the information about requirements. This is followed by a section in which I formulate those requirements in a more concrete way. The fourth section is the proposed solutions for the given requirements. The last section is the conclusions.

\section{Search for the requirements}
\subsection{General ethical considerations}
The issue of the ethics in the AI and data field is a field of study in its own right \cite{Kazim2021} as is the field of the medical ethics. I will refrain from the full discussion of these matters and focus on the issues from the general computer science ethics point of view. There is plenty of literature available about the ethics for AI in healthcare see for example references \cite{Char2020}, \cite{Geis2019}, \cite{Pesapane2018}, \cite{Miguel2020}, \cite{Mirbabaie2021}, \cite{Morley_2019}, and \cite{Morley2020}.

In the software development practically every design decision has to be justifiable after ethical analysis \cite{Kraemer2011}. Simplest question is about how resource intensive can we be, how much do we value the output accuracy versus the resource usage to get that accuracy? Each technological project should also undergo proper impact assessment \cite{Wright2011}. 

This analysis is based on the framework provided in the reference \cite{Wright2011} which provides a list of questions for consideration. I will go through most of them in order, but skip those which are trivially irrelevant to this work.

Questions about respect for the autonomy are mostly trivially irrelevant, except the case of curtailing personal freedom of movement. If a person is diagnosed with the COVID he will most likely be quarantined. This is justified due to the public health risk which the infected persons pose to the others.

In the area of the dignity the questions are mostly trivially irrelevant, as the goal is to provide a quicker and less invasive way of diagnosing the COVID from patients with a pneumonia. The patients would be in any case subjected to the chest X-rays and the diagnosis tool would be used to analyze those images.

Next section in the reference \cite{Wright2011} is about the informed consent. In the case of the data collection one should be using datasets only from respectable institutions which have followed the required ethical practices in their work. When the AI model would be used by the medical practitioners can only give their informed consent if the quality of the AI system is evaluated properly and it has an explainability functionality. This explainability is required by the medical practitioners to do informed decisions based on the output of the system \cite{Poon2021}. Another issue here is the collection on data during the use of the system.

Non-maleficence section of the reference \cite{Wright2011} starts with safety related questions. As this would be used as a clinical tool the safety aspect is a critical one. There will be a great harm coming from the errors in the diagnosis. A false positive diagnosis will cause psychological harm to the patient and a false negative will slow down the treatment and puts other people in risk as well. The algorithm needs a thorough testing before it can be used in a clinical settings.

Second part of this section in the reference \cite{Wright2011} is the social solidarity, and inclusion and exclusion. This is mainly about the information society inclusion and only relevant to this discussion is the fact that the system should be available for offline use in the areas where the internet connectivity is not good. 

The beneficence section of the reference \cite{Wright2011} has multiple questions which are relevant to this work. The goal is to benefit the individuals and the society by a faster and less invasive way to diagnose the COVID. With the X-ray image analysis the diagnosis can be done in 30 minutes \cite{Pan2021}, while the nasal swap and the RT-PCR will take at minimum multiple hours to complete \cite{Jawerth2020}. This should be a great benefit for all humans. 

Next section in the reference \cite{Wright2011} is the universal service. It should be available in all medical stations equipped with the X-ray machinery and a computer.  Here one should also take into account the issue of the global computer part shortage which limits the ability to get the newer and more powerful computing equipment \cite{Cooney2021},\cite{Economist2021}, \cite{Patrizio2021}, \cite{NE2021}. The diagnosis software thus should be usable in any modern computer. 
 
The accessibility is also an issue which needs some discussion. In the case of fully usable software an extremely simple user interface so that it is easy to use with a minimal training. If the goal is only produce the model for diagnostic, it has to have a simple and well documented API. 

The value sensitive design has some relevance here. The explainable nature of the AI will provide empowerment to the medical personnel. With a “regular” AI they will get only a value stating that the patient has the COVID with some probability, with an eXplainable AI they also get information on why the AI has come to this conclusion. This will provide them a lot more information and they can use it for their benefit.

For the sustainability we have some issues with possible change in the standards. The system should be built using the current standards and a modular design so that it could be easily updated.

The justice section in the reference \cite{Wright2011} discusses about distributive justice for all individuals and groups. This is highly relevant matter for any diagnosis tool as it is widely known that there are problems with many diagnosis methods when the patient is not a Caucasian male. See for example references \cite{Mahendraraj2017}, and \cite{Silveyra2021}. To overcome this problem the dataset needs to have data with an excellent reach over all humanity, not just in a single demographic group. The less desirable alternative solution is to state clearly the issues and limitations of the diagnosis model in the publication.

The equality and fairness (social justice) as defined in the reference \cite{Wright2011} is not as relevant as the previous part. The main point in it is the availability of the service for all, not just to a segment of the population based on their privileges. This is solved by the same solution as the universal service issues. Another point raised is the risk for the diagnosis being used for detriment of the patient. 

Next part in the reference \cite{Wright2011} talks about privacy and data protection. This is mainly relevant to the training data which I have discussed earlier. These issues is addressed when selecting the data source and by not collecting data during the use of the system.  

\subsection{General Data Science and AI related considerations}

Almost every AI project can be seen as a data science project in which we are tasked to find new knowledge from the data. For this there are well established workflows like KDD \cite{Fayyad1996} and CRISP-DM \cite{crisp}, and others for which one can see reference \cite{Mariscal2010} which provides a review.

Both the KDD and the CRISP-DM start with the understanding of the domain. One has to have enough domain knowledge to know what is needed and which things are relevant. This includes the knowledge on what has to be reported.

Next step is the data understanding and preparation. One has to know the data properly for identifying possible biases, confounding factors which could lead to shortcut learning \cite{Geirhos2020}, imbalance \cite{He2009} and other possible issues in the data quality and suitability for the task. Data selection is one of the most critical tasks.

This is followed by the modeling. This requires selection of suitable modeling tools for the data and the goal.

The statistical model analysis is then performed, which includes proper review of models properties related to the goal. Here we need to remember that any data which has been used in the model building cannot be used in the evaluation \cite{Cureton1950}, \cite{Kurtz1948} (including those which are not independent from the data used in the building.) One should remember that there are multiple ways of doing the estimation of the generalization performance and choose the most appropriate one \cite{Xu2018}.

Last step is deployment in to the production.

The usual structure of the project is also iterative one, so based on the discoveries in each step one goes back to a suitable step and acts upon the changed situation. Like if you find confounding factors then you go to the data preparation to remove them.  

For AI tasked to image analysis CNN \cite{Fukushima1980},\cite{Lecun1998} is the default choice of the architecture, but there are options like Capsule Neural Network (CapsNet) \cite{Sabour2017}, Graph Neural Network (GNN) \cite{Gori2005}, and their combination CapsGNN \cite{Xinyi2019}. One interesting approach is Multiple Instance Learning \cite{Keeler1991} and \cite{Dietterich1997}. These all require proper understanding of the data for the proper architecture implementation. The decision between different options depends on the task specific details and data availability.

\begin{figure*}[!t]
\centering
\resizebox{1.267\totalheight}{!}{
\subfloat[Original CXR]{\includegraphics[width=2.5in]{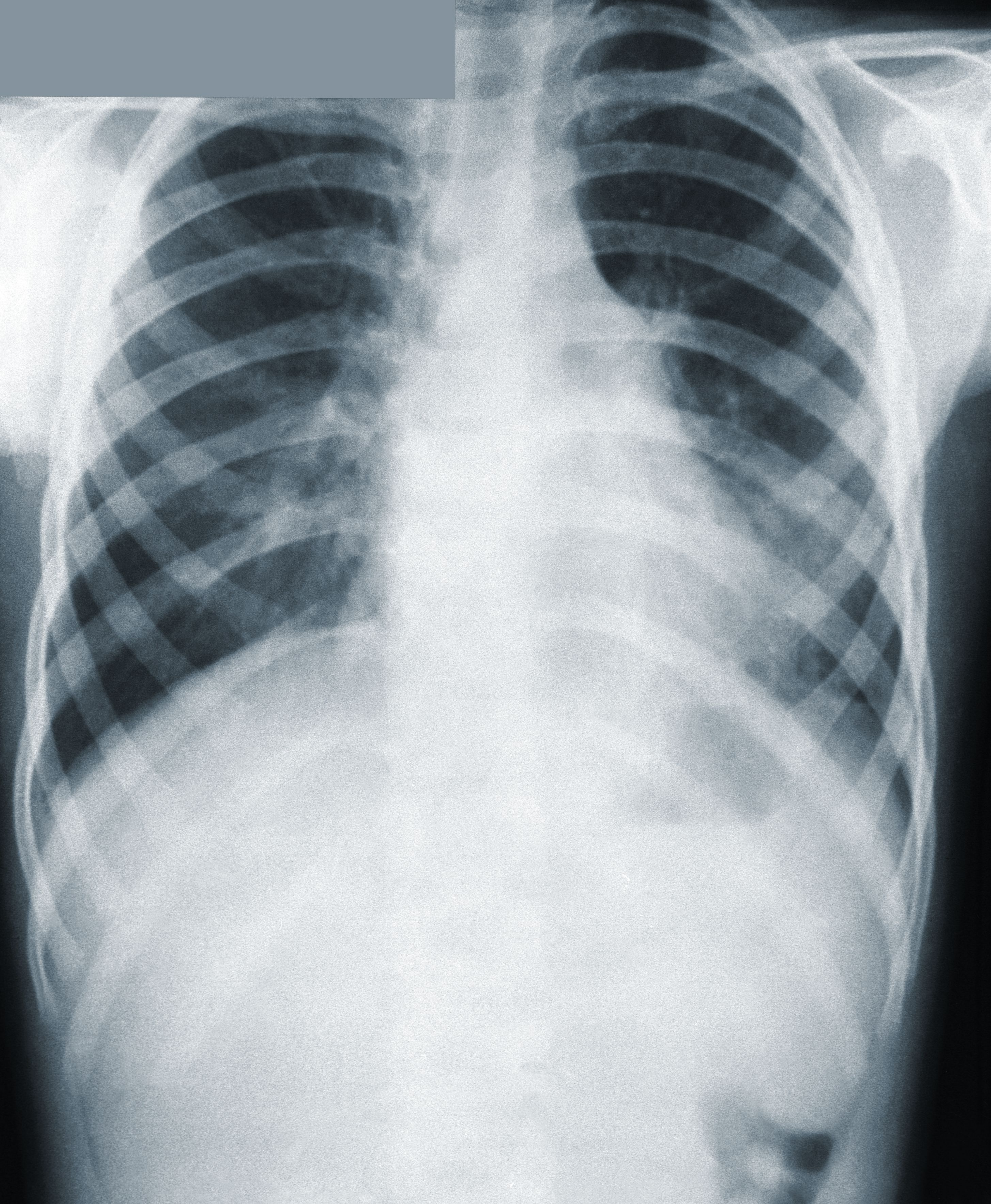}}
\hfil
\subfloat[Flipped CXR]{\includegraphics[width=2.5in]{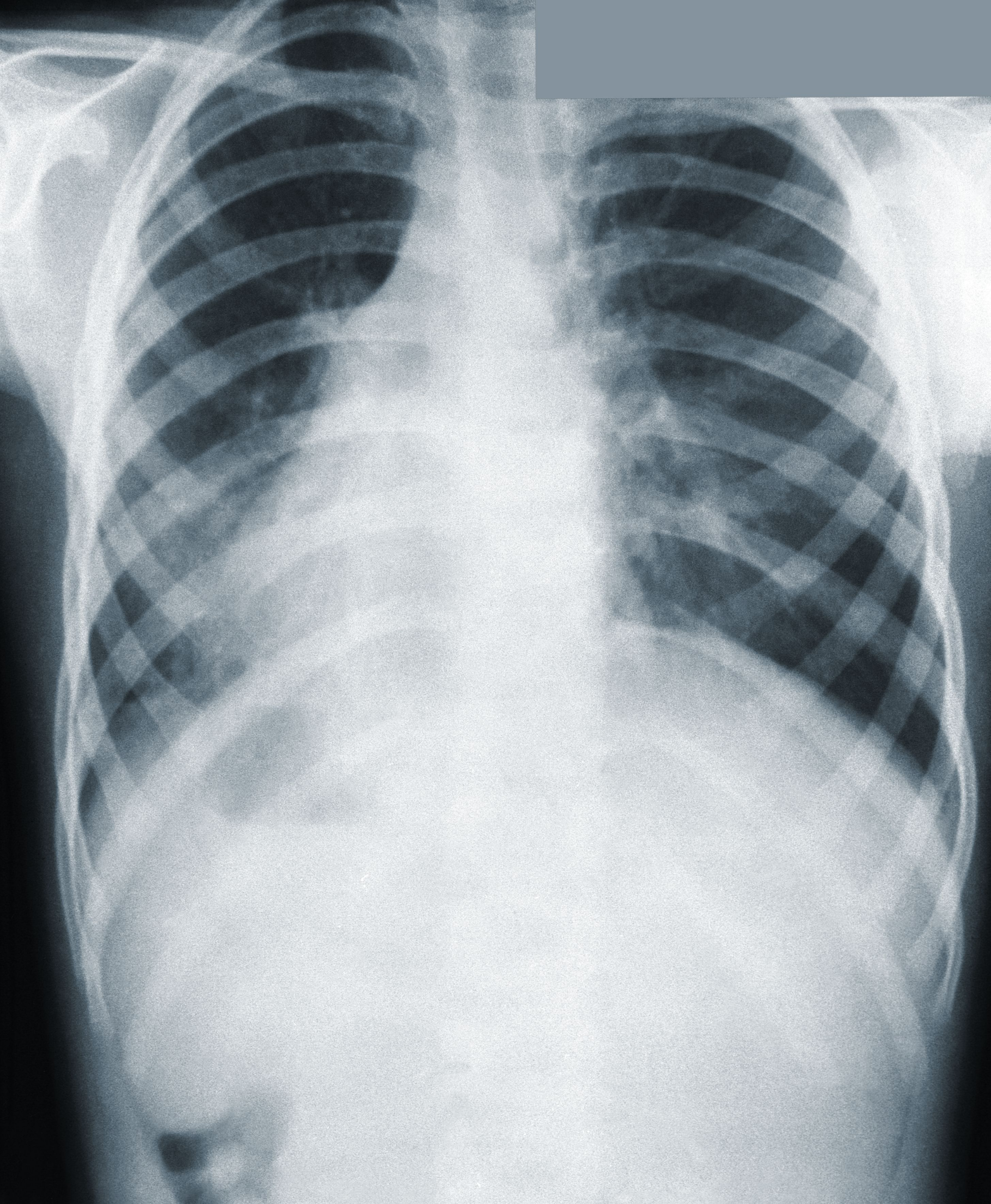}}
}
\caption{An example of incorrect augmentation, note how the internal organs are in incorrect position after flipping the image horizontally. Original photo is Public Domain \href{https://phil.cdc.gov/Details.aspx?pid=21493}{from CDC}.}
\label{fig:flip}
\end{figure*}

A bias in data, as an extreme example all in one group having notable but non-relevant feature while on the other group no-one has this feature, can influence how the model does the classification. There are options for handling data imbalance \cite{Sun2009}, \cite{Kim2019} and scarcity.Before using any of the standard image augmentation techniques \cite{Shijie2017} one must have enough domain knowledge to refrain generating incorrect data, as a simple example I use rotating 6 to 9 in number recognition. Example of this is found in the figure \ref{fig:flip}.
 
\subsection{The analysis of the reviews done to the previous COVID-19 diagnosis models}
There has been many papers published for reviewing work done to diagnose the COVID-19 from the CXR- and CT-images. Reviews presented in this section are focused on the clinical applicability of the reviewed models and thus are extremely critical as the field has extremely strict regulation and tight tolerances. I have used these reviews for gathering requirements.

The reference \cite{Burlacu2020} was among the first ones and it studied 14 publications of the AI models for the COVID-19 detection published by the end of March 2020. It found issues with testing for biases, the used datasets, and with used algorithms, including the use of out-of-the box models and lack of the explainability.

The use of different classification methods (binary, multiple classes, multiple labels, and hierarchical) in the COVID diagnosis was researched in the reference \cite{Albahri2020}. They found 11 studies related to their interest by May 5 2020. They found out that on those there were lack of consistency in the evaluating the quality of their model predictions.

Publications between March 2020 and May 2020 were also reviewed in the reference \cite{Alghamdi2021}. They found 34 publications and found that many publications had issues with dataset selection and possibly used multiple copies of same images. Other reported issues with used data were the class imbalances, private datasets. They also criticized the lack of uniformity in the quality evaluation, including bias evaluation. Explainability was also an issue which they brought up.
 
According to reference \cite{Roberts2021} studied papers proposing ML based diagnosis of the COVID-19 with data from the CXR or the CT scans from 1 January 2020 to 3 October 2020. They found 320 papers for their quality review and 258 failed in the first section of their analysis. Insufficient documentation in the model selection (132 failures), the methods of pre-processing of the images (125), and the details of the training approach (105) were the three most common failures. One critical failure was not disclosing the dataset used in the analysis. For the papers which passed this screening among the reasons for failing the clinical suitability were the lack of the proper validation, the robustness or sensitivity analysis, the demographics of the people in the data, the statistical testing for results, and the reporting issues regarding to the generalization. The paper criticized the lack of attention given to the features of the used datasets, for example using the dataset\cite{Kermany2018} as the control set while it consists of pediatric patients aged between on and five, and the COVID-19 patients were adults. Another issue raised was the downward scaling of the images due to the use of the ready of-the-shelf models. This and the lack of the demographic data is also related to the use of JPEG and PNG images instead of the DICOM \cite{dicom} which has metadata of the image acquisition parameters and other important information. The code availability and other replicability issues were also mentioned as was the call for the interpretability.

Another paper dedicated for creating proper basis for responsible deep learning for diagnosing COVID-19 from medical images is reference \cite{Hryniewska2021}. This paper is centered around use of the Explainable Artificial Intelligence to find the possible errors in the model, but was not limited to it. They analyzed 25 models, which were collected by August 14 2020. They found mistakes in data acquisition, model development, and explainability. As an example the paper mentioned that one should not use image augmentations which produce "impossible" images. For future models it produced a checklist for creating a responsible deep learning model for this task.

Reference \cite{Wynants2020} presents a systemic review of 169 studies. It provides a critical appraisal of the prediction models for the diagnosis and the prognosis of the COVID-19 in the selected studies which also included other implementations than the image analysis of CXR or CT. They found that all studies had either a high or an unknown risk of a bias in the results. This was mainly caused by non-representative selection of control patients, overfitting, issues with result validation, and unclear reporting.

Many other papers have raised the issue of lack of generalization of the AI diagnosis due to confounding factors in the data \cite{Maguolo2021}, \cite{Zech2018}, \cite{DeGrave2021}, \cite{Ahmed2021}, \cite{Pooch2020}, \cite{Tartaglione2020}. This leads to diagnosis based on other issues than the actual relevant features on the lung area, like learning from which data set the image is based on other information visible in the image. Another recognized source for lack of generalization was the dataset biases \cite{Janizek2020}, \cite{Candemir2019}, \cite{Saez2021}, \cite{Tartaglione2020}. Possible sources for these biases include patient demographics, procedures (for example the direction from which the X-ray image was taken), and procedures performed before taking the X-ray (for example an intubation tube visible in the CXR image.)

Reference \cite{Asgharnezhad2020} calls for uncertainty evaluation for predictions. This is related to lack of statistical analysis of the prediction quality mentioned in both references \cite{Hryniewska2021} and \cite{Roberts2021}.

\subsection{Analysis of other AI related medical publications}
There are papers written on the use of the AI for diagnosis or for other purposes in the medical science. In this section I will discuss the issues which they have raised.

The medical papers on the AI use outside of the COVID-19 are also relevant as they provide information on what is needed from a AI product used in the medicine. Reference \cite{Kourou2015} called for the good training data, the performance validation, and was critical of the "black-box" nature of some AI models. Papers like \cite{Choy2018}, \cite{Deo2015}, \cite{Lee2017},  \cite{Kallianos2019}, and \cite{Monshi2020} reminded on issues with the data and the "black-box". Reference \cite{Monshi2020} even called the "black-box" AI as unacceptable in the medical domain. The lack of the documentation for reproducibility was also brought up by the reference \cite{Kallianos2019}. Among other papers calling for the XAI are references \cite{Fellous2019}, \cite{He2019}, \cite{Holzinger2017}, \cite{Holzinger2019}, \cite{Litjens2017}, and \cite{Yu2018}. The reference \cite{Altaf2019} points out issues with amount of data available for training and imbalance issues in the data, and it also point out the lack of the confidence intervals in the predictions.

The Checklist for Artificial Intelligence in Medical Imaging (CLAIM) is available in the reference \cite{Mongan2020}. There is also a radiomics quality score (RQS) \cite{Lambin2017}, which can be used.

Transparent reporting of a multivariable prediction model for individual prognosis or diagnosis (TRIPOD)\cite{Collins2015} gives a reporting guidelines for diagnosis tools.

The PROBAST (Prediction model Risk Of Bias ASsessment Tool)\cite{Wolff2019} tool is commonly used for the estimation of the risk of bias. The bias is usually related to the dataset not being representative due to the data scarcity, the population shift, the prevalence shift, or the selection bias \cite{Castro2020}, \cite{Almasni2020}. The dataset shifts are also discussed in reference \cite{Subbaswamy2019}. 

The reference \cite{Badgeley2019} noted confounding factors in the medical image analysis of the knee is also.

The reference \cite{Oala2020} describes the application of the ITU/WHO FG-AI4H (Focus Group on Artificial Intelligence for Health) assessment guidelines \cite{fg-ai4h} for a machine learning tool. The application includes a questionnaire about the project, the bias \& fairness analysis, the interpretability (explainability), the robustness evaluation and the reporting guidelines.

\section{Requirements found}
\subsection{Ethical requirements}
In this section I will transfer the thoughts presented in the ethical considerations section into more proper requirements.

\begin{enumerate}
\item The data collection has to be done with an informed consent or use data from a collection by a respectable source.
\item The quality of the diagnosis has to be known to the medical practitioners.
\item Medical practitioners needs to have explanation why the software made the diagnosis.
\item Product needs to be usable in a low infrastructure area. This means that the software should be able to be used without an internet connection and with a low end computer.
\item Faster or better than the competing technologies (see for example the references \cite{Pan2021}, \cite{Jawerth2020}, and \cite{Leli2021}). The diagnosis quality better than rapid antigen testing and result should be available in less than 10 minutes with an standard medium or low power computer.
\item Good and simple user interface. It can be limited to a simple and well documented API.
\item Use eXplainable AI to empower the users.
\item Use medical imaging standards (DICOM\cite{dicom}).
\item The model should work as well with all segments of the human population. If this is not possible it should be clearly stated.
\item Do not collect data during the operation without consent.
\end{enumerate}

\subsection{General data science and AI requirements}

Here we basically have only two main requirements.

\begin{enumerate}
\item Learn the domain issues.
\item Study the data properly to find any possible issues.
\end{enumerate}

while others are
\begin{enumerate}
\setcounter{enumi}{2}
\item Select only proper data.
\item Fix any issues with the data.
\item Select suitable modeling tools for the data.
\item Perform proper analysis of the model created.
\item Repeat if result is not good enough.
\end{enumerate}

\subsection{Clinical AI requirements for the COVID diagnosis}
In order to generalize the results (or at least know the limits of the generalization) we need to:
\begin{enumerate}
\item Remove the confounding factors.
\item Handle the biases in the datasets.
\item Handle the data scarcity.
\item Handle the shifts between the datasets and the reality.
\end{enumerate}

For reliability review which is needed for every medical application we need to:
\begin{enumerate}
\setcounter{enumi}{4}
\item Do proper the documentation of choices made.
\item Do the result validation.
\item Do the robustness and sensitivity analysis.
\item Select the image size for the analysis based on scientific reasons, not based on the convenience.
\item Use the metadata available.
\item Have the code and other material available for replication.
\item Explain the reasons for given diagnosis.
\end{enumerate}

\subsection{Requirements from the other medical AI publications}

Here we have following:
\begin{enumerate}
\item Use checklist(s).
\item Explainability.
\item Bias analysis.
\item Proper data issue handling.
\item Documentation.
\item Performance validation.
\end{enumerate}

\subsection{Combined requirements}
Many of these requirements are overlapping and thus the list of the requirements can be simplified to following.

\begin{itemize}
\item Learn the basics of the medical imaging domain.
\item Use checklist(s).
\item Use the DICOM data from a respectable source.
\item Study the available data properly.
\item Handle data issues: selection, biases, confounding factors, scarcity, and shifts.
\item Select a suitable model with explainability.
\item Document every decision, and the reasons for them, regarding to the data, the model architecture, and the (meta)parameters of the model.
\item Do the proper statistical analysis of the quality of the model predictions.
\item The model needs to be better than other diagnosis methods. (Faster, more accurate, or better availability in low infrastructure areas.)
\item Store everything needed for the replication of the results.
\end{itemize}

\section{Practical solutions to these requirements}

The requirements and solutions listed here are aimed to be used for the preliminary studies to find suitable models. For the proper handling the issues do take a look at the references given here and previously in this work and legal requirements for clinical applications. My recommendations are also found in a condensed form in the table \ref{tab1}.

\subsection{Learn the basics of the medical imaging domain.}
The best solution here is to include a domain expert into the group, the absolute minimum is a proper review of the previous work done in the field.

\subsection{Use checklists}
There are multiple checklists e.g references \cite{Hryniewska2021}, \cite{Mongan2020}, \cite{Wolff2019}, \cite{fg-ai4h}, \cite{Lambin2017}, \cite{Collins2015}

\subsection{Use full sized DICOM data from a respectable source}
This has its own solution written directly in to the requirement. As DICOM is an industry standard it is well documented and there exists ready libraries for its use. The respectable source would be some proper institute, not a private collection.

\subsection{Study the available data properly.}
Do proper analysis to the raw data. This has to include study of demographics, biases, duplicates, outliers, pixel intensities, etc.

\subsection{Handle data issues: Selection, biases, confounding factors, scarcity, and shifts.}
In the data selection one needs to be careful and one has to remove the confounding factors and the incorrect data. Especially the area outside of lungs holds many confounding artifacts \cite{Hryniewska2021}, while the secondary source of confounding is the pixel intensity \cite{Pooch2020}. There are plenty of ready lung segmentation models available, but they should be reviewed properly before selection. Incorrect data should not be included, examples of this are duplicates and failed images.

In the case of the bias this can sometimes be handled with the proper selection of the used data or with generating new data via the augmentation\cite{Shorten2019}, \cite{Stephen2019} or via the generative adversarial nets (GAN) \cite{Goodfellow2014}. The last resort is to record properly the existing biases and continue with them. On the new data generation one needs to be extremely careful not to use incorrect techniques \cite{Hryniewska2021}.

For data scarcity we have a possibility to generate new data as in the case of bias and other solution is the transfer learning \cite{Weiss2016}, \cite{Tan2018}, \cite{Abbas2020}, \cite{Karimi2021}. But one should remember that transfer learning is not always useful \cite{Raghu2019}, \cite{Cheplygina2019}.

Dataset shifts are discussed with detail in the references \cite{Castro2020} and \cite{Subbaswamy2019}.

\subsection{Select a suitable model with explainability.}
CNN is still the default choice, but if other relevant factors points toward choosing something else do not discard them. The available out of the box models are not always the best choice \cite{Raghu2019}, \cite{Stephen2019}.

There are multiple different explanation tools for the image classification AI. The reference \cite{Hryniewska2021} gives some overview what has been used previously and gives some pointers on the issues related to them. The suitability of these explanation tools for other than the CNN is also an issue. 

\subsection{Documentation}
Document every decision, and the reasons for them, regarding to the data, the model architecture, and the (meta)parameters of the model.

\subsection{Do a proper statistical analysis of the quality of the model predictions.}
According to the reference \cite{Albahri2020} we need following statistical information. 
\begin{description}
\item [Binary categorization]  \hfill \\ 
 Accuracy, precision, recall (sensitivity), F score, specificity, AUC.
\item [Multi-class classification]  \hfill \\ 
 Average accuracy, error rate, precision$_\mu$, recall$_\mu$, F score$_\mu$, precision$_M$, recall$_M$, F score$_M$ 
\item [Multi-label classification]  \hfill \\ 
 Exact match ratio, labeling F score, retrieval F score, Hamming loss.
\item [Hierarchical classification] \hfill \\
 Precision$\downarrow$, recall$\downarrow$, F score$\downarrow$, precision$\uparrow$, recall$\uparrow$ and F score$\uparrow$, which are defined in the reference \cite{Albahri2020}.
\end{description}

\subsection{Model performance}
Compare the model to the current state of the art with other techniques like the RT-PCR. Remember to use up to date information on this comparison.

\subsection{Store everything needed for replication of the results}
Use the GitHub or some other similar service.

\begin{table*}[th]
\caption{Requirements and proposed solutions}
\begin{center}
\begin{tabular}{|c|c|}
\hline
\textbf{Requirement} & \textbf{Solution}\\
\hline
\hline
Learn the domain & Get a domain expert or read the proper source papers.\\
\hline
Use checklists & Lists are found in references \cite{Hryniewska2021}, \cite{Mongan2020}, \cite{Lambin2017}, \cite{Collins2015}, \cite{Wolff2019} \\
\hline
Data gathering & Use the DICOM data from a respectable sources\\
\hline
Data preview & Do a proper preview of the data properties to find the possible issues.\\
\hline
Data issues & Handle the issues with care, see at least the references \cite{Hryniewska2021}, \cite{Castro2020}, and \cite{Subbaswamy2019}.\\
\hline
Select model & Find the proper model suitable for the task and equip it with the explanability module.\\
\hline
Documentation & Document and explain all choices made to the regarding data, the architecture, and the (meta)parameters.\\
\hline
Statistical analysis & Follow the reference \cite{Albahri2020}\\
\hline
Model performance & Compare to other technologies like the RT-PCR\\
\hline
Replication & Store everything needed for it in the GitHub or in an other similar service.\\
\hline
\end{tabular}
\label{tab1}
\end{center}
\end{table*}

\section{Conclusions}
There has been a great effort and lots of enthusiasm for providing an AI solution to the clinical diagnosis of the COVID using the CXR data. While promising results have been published, unfortunately most of the publications lack the rigor needed in the medical field. This issue is shown by multiple reviews \cite{Hryniewska2021}, \cite{Roberts2021}, \cite{Wynants2020}, \cite{Burlacu2020}, \cite{Albahri2020}, and \cite{Alghamdi2021} and our field needs to pay a proper respect to the actual requirements for such tools. There are two major limitations in this work, the lack of proper medical expertise by the author and reliance to previous review papers which all always behind the state of the art implementations which have been published after articles for reviews are selected.

This work is a start into this direction and a pointer towards more thorough work done by the domain experts. This work can also be used as a basis for other analyses on similar issues. Personally this is the starting point for developing new AI model for the diagnosis of the COVID-19 from chest X-ray images.

\bibliographystyle{IEEEtran}
\bibliography{IEEEabrv,requirements_workshop_paper_accepted.bib}

\begin{thebibliography}{10}
\providecommand{\url}[1]{#1}
\csname url@samestyle\endcsname
\providecommand{\newblock}{\relax}
\providecommand{\bibinfo}[2]{#2}
\providecommand{\BIBentrySTDinterwordspacing}{\spaceskip=0pt\relax}
\providecommand{\BIBentryALTinterwordstretchfactor}{4}
\providecommand{\BIBentryALTinterwordspacing}{\spaceskip=\fontdimen2\font plus
\BIBentryALTinterwordstretchfactor\fontdimen3\font minus
  \fontdimen4\font\relax}
\providecommand{\BIBforeignlanguage}[2]{{%
\expandafter\ifx\csname l@#1\endcsname\relax
\typeout{** WARNING: IEEEtran.bst: No hyphenation pattern has been}%
\typeout{** loaded for the language `#1'. Using the pattern for}%
\typeout{** the default language instead.}%
\else
\language=\csname l@#1\endcsname
\fi
#2}}
\providecommand{\BIBdecl}{\relax}
\BIBdecl

\bibitem{WHO_PHEIC}
\BIBentryALTinterwordspacing
{World Health Organization}, ``Novel coronavirus (2019-{nCoV}): situation
  report, 11,'' World Health Organization, Technical documents, 2020-01-31.
  [Online]. Available: \url{https://apps.who.int/iris/handle/10665/330776}
\BIBentrySTDinterwordspacing

\bibitem{WHO_pandemic}
\BIBentryALTinterwordspacing
------, ``Coronavirus disease 2019 ({COVID}-19): situation report, 51,'' World
  Health Organization, Technical documents, 2020-03-11. [Online]. Available:
  \url{https://apps.who.int/iris/handle/10665/331475}
\BIBentrySTDinterwordspacing

\bibitem{Hryniewska2021}
W.~Hryniewska, P.~Bombiński, P.~Szatkowski, P.~Tomaszewska, A.~Przelaskowski,
  and P.~Biecek, ``Checklist for responsible deep learning modeling of medical
  images based on {COVID}-19 detection studies,'' \emph{Pattern Recognition},
  vol. 118, p. 108035, 2021.

\bibitem{Roberts2021}
M.~Roberts, D.~Driggs, M.~Thorpe, J.~Gilbey, M.~Yeung, S.~Ursprung
  \emph{et~al.}, ``Common pitfalls and recommendations for using machine
  learning to detect and prognosticate for {COVID}-19 using chest radiographs
  and {CT} scans,'' \emph{Nature Machine Intelligence}, vol.~3, no.~3, pp.
  199--217, 2021.

\bibitem{Wynants2020}
L.~Wynants, B.~Van~Calster, G.~S. Collins, R.~D. Riley, G.~Heinze, E.~Schuit
  \emph{et~al.}, ``Prediction models for diagnosis and prognosis of covid-19:
  systematic review and critical appraisal,'' \emph{BMJ}, vol. 369, 2020.

\bibitem{Burlacu2020}
A.~Burlacu, R.~Crisan-Dabija, I.~V. Popa, B.~Artene, V.~Birzu, M.~Pricop
  \emph{et~al.}, ``Curbing the {AI}-induced enthusiasm in diagnosing {COVID}-19
  on chest {X}-rays: the present and the near-future,'' \emph{medRxiv}, 2020.

\bibitem{Albahri2020}
O.~S. Albahri, A.~A. Zaidan, A.~S. Albahri, B.~B. Zaidan, K.~H. Abdulkareem,
  Z.~T. Al-qaysi \emph{et~al.}, ``Systematic review of artificial intelligence
  techniques in the detection and classification of {COVID}-19 medical images
  in terms of evaluation and benchmarking: Taxonomy analysis, challenges,
  future solutions and methodological aspects,'' \emph{Journal of Infection and
  Public Health}, vol.~13, no.~10, pp. 1381--1396, 2020.

\bibitem{Alghamdi2021}
H.~S. Alghamdi, G.~Amoudi, S.~Elhag, K.~Saeedi, and J.~Nasser, ``Deep learning
  approaches for detecting {COVID}-19 from chest {X}-ray images: A survey,''
  \emph{IEEE Access}, vol.~9, pp. 20\,235--20\,254, 2021.

\bibitem{Kazim2021}
E.~Kazim and A.~S. Koshiyama, ``A high-level overview of {AI} ethics,''
  \emph{Patterns}, vol.~2, no.~9, p. 100314, 2021.

\bibitem{Char2020}
D.~S. Char, M.~D. Abràmoff, and C.~Feudtner, ``Identifying ethical
  considerations for machine learning healthcare applications,'' \emph{The
  American Journal of Bioethics}, vol.~20, no.~11, pp. 7--17, 2020, pMID:
  33103967.

\bibitem{Geis2019}
J.~R. Geis, A.~P. Brady, C.~C. Wu, J.~Spencer, E.~Ranschaert, J.~L. Jaremko
  \emph{et~al.}, ``Ethics of artificial intelligence in radiology: Summary of
  the joint {E}uropean and {N}orth {A}merican multisociety statement,''
  \emph{Radiology}, vol. 293, no.~2, pp. 436--440, 2019, pMID: 31573399.

\bibitem{Pesapane2018}
\BIBentryALTinterwordspacing
F.~Pesapane, C.~Volonté, M.~Codari, and F.~Sardanelli, ``Artificial
  intelligence as a medical device in radiology: ethical and regulatory issues
  in {E}urope and the {U}nited {S}tates,'' \emph{Insights into Imaging},
  vol.~9, no.~5, pp. 745--753, 2018. [Online]. Available:
  \url{https://doi.org/10.1007/s13244-018-0645-y}
\BIBentrySTDinterwordspacing

\bibitem{Miguel2020}
I.~de~Miguel, B.~Sanz, and G.~Lazcoz, ``Machine learning in the {EU} health
  care context: exploring the ethical, legal and social issues,''
  \emph{Information, Communication \& Society}, vol.~23, no.~8, pp. 1139--1153,
  2020.

\bibitem{Mirbabaie2021}
\BIBentryALTinterwordspacing
M.~Mirbabaie, L.~Hofeditz, N.~R.~J. Frick, and S.~Stieglitz, ``Artificial
  intelligence in hospitals: providing a status quo of ethical considerations
  in academia to guide future research,'' \emph{AI \& SOCIETY}, 2021. [Online].
  Available: \url{https://doi.org/10.1007/s00146-021-01239-4}
\BIBentrySTDinterwordspacing

\bibitem{Morley_2019}
J.~Morley, C.~Machado, C.~Burr, J.~Cowls, M.~Taddeo, and L.~Floridi, ``The
  debate on the ethics of {AI} in health care: a reconstruction and critical
  review,'' \emph{{SSRN} Electronic Journal}, 2019.

\bibitem{Morley2020}
J.~Morley, C.~C. Machado, C.~Burr, J.~Cowls, I.~Joshi, M.~Taddeo \emph{et~al.},
  ``The ethics of {AI} in health care: A mapping review,'' \emph{Social Science
  \& Medicine}, vol. 260, p. 113172, 2020.

\bibitem{Kraemer2011}
F.~Kraemer, K.~van Overveld, and M.~Peterson, ``Is there an ethics of
  algorithms?'' \emph{Ethics and Information Technology}, vol.~13, no.~3, pp.
  251--260, 2011.

\bibitem{Wright2011}
D.~Wright, ``A framework for the ethical impact assessment of information
  technology,'' \emph{Ethics and Information Technology}, vol.~13, no.~3, pp.
  199--226, 2011.

\bibitem{Poon2021}
A.~I.~F. Poon and J.~J.~Y. Sung, ``Opening the black box of {AI}-medicine,''
  \emph{Journal of Gastroenterology and Hepatology}, vol.~36, no.~3, pp.
  581--584, 2021.

\bibitem{Pan2021}
L.~Pan, J.~Zeng, H.~Pu, and S.~Peng, ``How to optimize the radiology protocol
  during the global {COVID}-19 epidemic: Keypoints from {S}ichuan provincial
  people's hospital,'' \emph{Clinical Imaging}, vol.~69, pp. 324--327, 2021.

\bibitem{Jawerth2020}
\BIBentryALTinterwordspacing
N.~Jawerth, ``How is the {COVID}-19 virus detected using real time {RT-PCR}?''
  \emph{IAEA Bulletin (Online)}, vol.~61, no.~2, pp. 8--11, 2020. [Online].
  Available: \url{https://www.iaea.org/sites/default/files/6120811.pdf}
\BIBentrySTDinterwordspacing

\bibitem{Cooney2021}
\BIBentryALTinterwordspacing
M.~Cooney, ``Chip shortage will hit {IT}-hardware buyers for months to years:
  Tech executives and analysts say the current processor-chip shortage and
  disruption of supply chains thanks to {COVID}-19 could have a long-term
  impact on price and availability,'' \emph{Network World (Online)}, 2021.
  [Online]. Available:
  \url{https://www.networkworld.com/article/3619210/chip-shortage-will-hit-it-hardware-buyers-for-months-to-years.html}
\BIBentrySTDinterwordspacing

\bibitem{Economist2021}
\BIBentryALTinterwordspacing
{The Economist Intelligence Unit}, ``The global chip shortage is here for some
  time: Loading, please wait,'' \emph{Global Business Review}, 2021. [Online].
  Available:
  \url{https://www.economist.com/finance-and-economics/2021/05/20/the-global-chip-shortage-is-here-for-some-time}
\BIBentrySTDinterwordspacing

\bibitem{Patrizio2021}
\BIBentryALTinterwordspacing
A.~Patrizio, ``You’re not imaging things, there is a serious chip shortage:
  {CPUs}, {GPUs}, and memory are all in tight supply due to manufacturing
  issues and high demand,'' \emph{Network World (Online)}, 2021. [Online].
  Available:
  \url{https://www.networkworld.com/article/3623753/the-chip-shortage-is-real-but-driven-by-more-than-covid.html}
\BIBentrySTDinterwordspacing

\bibitem{NE2021}
\BIBentryALTinterwordspacing
``Chips in a crisis,'' \emph{Nature Electronics}, vol.~4, no.~5, pp. 317--317,
  2021. [Online]. Available: \url{https://doi.org/10.1038/s41928-021-00601-0}
\BIBentrySTDinterwordspacing

\bibitem{Mahendraraj2017}
K.~Mahendraraj, K.~Sidhu, C.~S.~M. Lau, G.~J. McRoy, R.~S. Chamberlain, and
  F.~O. Smith, ``Malignant melanoma in {A}frican-{A}mericans: A
  population-based clinical outcomes study involving 1106 {A}frican-{A}merican
  patients from the surveillance, epidemiology, and end result ({SEER})
  database (1988-2011).'' \emph{Medicine}, vol.~96, p. e6258, 2017.

\bibitem{Silveyra2021}
P.~Silveyra and X.~Tigno, Eds., \emph{Sex-Based Differences in Lung
  Physiology}.\hskip 1em plus 0.5em minus 0.4em\relax Springer International
  Publishing, 2021.

\bibitem{Fayyad1996}
U.~M. Fayyad, G.~Piatetsky-Shapiro, and P.~Smyth, \emph{From Data Mining to
  Knowledge Discovery: An Overview}.\hskip 1em plus 0.5em minus 0.4em\relax
  USA: American Association for Artificial Intelligence, 1996, p. 1–34.

\bibitem{crisp}
\BIBentryALTinterwordspacing
P.~Chapman, J.~Clinton, R.~Kerber, T.~Khabaza, T.~Reinartz, C.~Shearer
  \emph{et~al.}, ``{CRISP-DM} 1.0 step-by-step data mining guide,'' The
  CRISP-DM consortium, Tech. Rep., August 2000. [Online]. Available:
  \url{https://maestria-datamining-2010.googlecode.com/svn-history/r282/trunk/dmct-teorica/tp1/CRISPWP-0800.pdf}
\BIBentrySTDinterwordspacing

\bibitem{Mariscal2010}
G.~Mariscal, {\'{O}}.~Marb{\'{a}}n, and C.~Fern{\'{a}}ndez,
  ``\BIBforeignlanguage{English}{A survey of data mining and knowledge
  discovery process models and methodologies},''
  \emph{\BIBforeignlanguage{English}{The Knowledge Engineering Review}},
  vol.~25, no.~2, pp. 137--166, 06 2010.

\bibitem{Geirhos2020}
R.~Geirhos, J.-H. Jacobsen, C.~Michaelis, R.~Zemel, W.~Brendel, M.~Bethge
  \emph{et~al.}, ``Shortcut learning in deep neural networks,'' \emph{Nature
  Machine Intelligence}, vol.~2, no.~11, pp. 665--673, 2020.

\bibitem{He2009}
H.~He and E.~A. Garcia, ``Learning from imbalanced data,'' \emph{IEEE
  Transactions on Knowledge and Data Engineering}, vol.~21, no.~9, pp.
  1263--1284, Sep. 2009.

\bibitem{Cureton1950}
\BIBentryALTinterwordspacing
E.~E. Cureton, ``{V}alidity, reliability, and baloney,'' \emph{Educational and
  Psychological Measurement}, vol.~10, no.~1, pp. 94--96, 1950. [Online].
  Available: \url{https://doi.org/10.1177/001316445001000107}
\BIBentrySTDinterwordspacing

\bibitem{Kurtz1948}
\BIBentryALTinterwordspacing
A.~K. Kurtz, ``A research test of the {R}orschach test,'' \emph{Personnel
  Psychology}, vol.~1, no.~1, pp. 41--51, 1948. [Online]. Available:
  \url{https://onlinelibrary.wiley.com/doi/abs/10.1111/j.1744-6570.1948.tb01292.x}
\BIBentrySTDinterwordspacing

\bibitem{Xu2018}
\BIBentryALTinterwordspacing
Y.~Xu and R.~Goodacre, ``On splitting training and validation set: A
  comparative study of cross-validation, bootstrap and systematic sampling for
  estimating the generalization performance of supervised learning,''
  \emph{Journal of Analysis and Testing}, vol.~2, no.~3, pp. 249--262, 2018.
  [Online]. Available: \url{https://doi.org/10.1007/s41664-018-0068-2}
\BIBentrySTDinterwordspacing

\bibitem{Fukushima1980}
\BIBentryALTinterwordspacing
K.~Fukushima, ``Neocognitron: A self-organizing neural network model for a
  mechanism of pattern recognition unaffected by shift in position,''
  \emph{Biological Cybernetics}, vol.~36, no.~4, pp. 193--202, 1980. [Online].
  Available: \url{https://doi.org/10.1007/BF00344251}
\BIBentrySTDinterwordspacing

\bibitem{Lecun1998}
Y.~Lecun, L.~Bottou, Y.~Bengio, and P.~Haffner, ``Gradient-based learning
  applied to document recognition,'' \emph{Proceedings of the IEEE}, vol.~86,
  no.~11, pp. 2278--2324, 1998.

\bibitem{Sabour2017}
S.~Sabour, N.~Frosst, and G.~E. Hinton, ``Dynamic routing between capsules,''
  in \emph{Proceedings of the 31st International Conference on Neural
  Information Processing Systems}, ser. NIPS'17.\hskip 1em plus 0.5em minus
  0.4em\relax Red Hook, NY, USA: Curran Associates Inc., 2017, p. 3859–3869.

\bibitem{Gori2005}
M.~Gori, G.~Monfardini, and F.~Scarselli, ``A new model for earning in raph
  domains,'' vol.~2, 01 2005, pp. 729 -- 734 vol. 2.

\bibitem{Xinyi2019}
\BIBentryALTinterwordspacing
Z.~Xinyi and L.~Chen, ``Capsule graph neural network,'' in \emph{International
  Conference on Learning Representations}, 2019. [Online]. Available:
  \url{https://openreview.net/forum?id=Byl8BnRcYm}
\BIBentrySTDinterwordspacing

\bibitem{Keeler1991}
\BIBentryALTinterwordspacing
J.~Keeler, D.~Rumelhart, and W.~Leow, ``Integrated segmentation and recognition
  of hand-printed numerals,'' in \emph{Advances in Neural Information
  Processing Systems}, R.~P. Lippmann, J.~Moody, and D.~Touretzky, Eds.,
  vol.~3.\hskip 1em plus 0.5em minus 0.4em\relax Morgan-Kaufmann, 1991.
  [Online]. Available:
  \url{https://proceedings.neurips.cc/paper/1990/file/e46de7e1bcaaced9a54f1e9d0d2f800d-Paper.pdf}
\BIBentrySTDinterwordspacing

\bibitem{Dietterich1997}
\BIBentryALTinterwordspacing
T.~G. Dietterich, R.~H. Lathrop, and T.~Lozano-Pérez, ``Solving the multiple
  instance problem with axis-parallel rectangles,'' \emph{Artificial
  Intelligence}, vol.~89, no.~1, pp. 31--71, 1997. [Online]. Available:
  \url{https://www.sciencedirect.com/science/article/pii/S0004370296000343}
\BIBentrySTDinterwordspacing

\bibitem{Sun2009}
\BIBentryALTinterwordspacing
Y.~Sun, A.~K.~C. Wong, and M.~S. Kamel, ``Classification of imbalanced data: A
  review,'' \emph{International Journal of Pattern Recognition and Artificial
  Intelligence}, vol.~23, no.~04, pp. 687--719, 2009. [Online]. Available:
  \url{https://doi.org/10.1142/S0218001409007326}
\BIBentrySTDinterwordspacing

\bibitem{Kim2019}
B.~Kim, H.~Kim, K.~Kim, S.~Kim, and J.~Kim, ``Learning not to learn: Training
  deep neural networks with biased data,'' in \emph{2019 IEEE/CVF Conference on
  Computer Vision and Pattern Recognition (CVPR)}, 2019, pp. 9004--9012.

\bibitem{Shijie2017}
J.~Shijie, W.~Ping, J.~Peiyi, and H.~Siping, ``Research on data augmentation
  for image classification based on convolution neural networks,'' in
  \emph{2017 Chinese Automation Congress (CAC)}, Oct 2017, pp. 4165--4170.

\bibitem{Kermany2018}
D.~S. Kermany, M.~Goldbaum, W.~Cai, C.~C.~S. Valentim, H.~Liang, S.~L. Baxter
  \emph{et~al.}, ``Identifying medical diagnoses and treatable diseases by
  image-based deep learning,'' \emph{Cell}, vol. 172, no.~5, pp. 1122--1131.e9,
  2018.

\bibitem{dicom}
\BIBentryALTinterwordspacing
\emph{{NEMA} {PS3} / {ISO} 12052, Digital Imaging and Communications in
  Medicine ({DICOM}) Standard}, National Electrical Manufacturers Association
  Std., Rev. 2021d, 2021. [Online]. Available:
  \url{http://www.dicomstandard.org}
\BIBentrySTDinterwordspacing

\bibitem{Maguolo2021}
G.~Maguolo and L.~Nanni, ``A critic evaluation of methods for {COVID}-19
  automatic detection from {X}-ray images,'' \emph{Information Fusion},
  vol.~76, pp. 1--7, 2021.

\bibitem{Zech2018}
J.~R. Zech, M.~A. Badgeley, M.~Liu, A.~B. Costa, J.~J. Titano, and E.~K.
  Oermann, ``Variable generalization performance of a deep learning model to
  detect pneumonia in chest radiographs: A cross-sectional study,'' \emph{PLOS
  Medicine}, vol.~15, no.~11, pp. 1--17, 2018.

\bibitem{DeGrave2021}
A.~J. DeGrave, J.~D. Janizek, and S.-I. Lee, ``{AI} for radiographic {COVID}-19
  detection selects shortcuts over signal,'' \emph{Nature Machine
  Intelligence}, 2021.

\bibitem{Ahmed2021}
K.~B. Ahmed, G.~M. Goldgof, R.~Paul, D.~B. Goldgof, and L.~O. Hall, ``Discovery
  of a generalization gap of convolutional neural networks on {COVID}-19
  {X}-rays classification,'' \emph{IEEE Access}, vol.~9, pp. 72\,970--72\,979,
  2021.

\bibitem{Pooch2020}
E.~H.~P. Pooch, P.~Ballester, and R.~C. Barros, ``Can we trust deep learning
  based diagnosis? the impact of domain shift in chest radiograph
  classification,'' in \emph{Thoracic Image Analysis}, J.~Petersen, R.~San
  Jos{\'e}~Est{\'e}par, A.~Schmidt-Richberg, S.~Gerard, B.~Lassen-Schmidt,
  C.~Jacobs \emph{et~al.}, Eds.\hskip 1em plus 0.5em minus 0.4em\relax Cham:
  Springer International Publishing, 2020, pp. 74--83.

\bibitem{Tartaglione2020}
E.~Tartaglione, C.~A. Barbano, C.~Berzovini, M.~Calandri, and M.~Grangetto,
  ``Unveiling {COVID}-19 from {CHEST} {X}-ray with deep learning: A hurdles
  race with small data,'' \emph{International Journal of Environmental Research
  and Public Health}, vol.~17, no.~18, 2020.

\bibitem{Janizek2020}
J.~D. Janizek, G.~Erion, A.~J. DeGrave, and S.-I. Lee, ``An adversarial
  approach for the robust classification of pneumonia from chest radiographs,''
  in \emph{Proceedings of the ACM Conference on Health, Inference, and
  Learning}, ser. CHIL '20.\hskip 1em plus 0.5em minus 0.4em\relax New York,
  NY, USA: Association for Computing Machinery, 2020, p. 69–79.

\bibitem{Candemir2019}
S.~Candemir and S.~Antani, ``A review on lung boundary detection in chest
  {X}-rays,'' \emph{International Journal of Computer Assisted Radiology and
  Surgery}, vol.~14, no.~4, pp. 563--576, 2019.

\bibitem{Saez2021}
C.~Sáez, N.~Romero, J.~A. Conejero, and J.~M. García-Gómez, ``Potential
  limitations in {COVID}-19 machine learning due to data source variability: A
  case study in the {nCov2019} dataset,'' \emph{J Am Med Inform Assoc},
  vol.~28, no.~2, pp. 360--364, 2021.

\bibitem{Asgharnezhad2020}
H.~Asgharnezhad, A.~Shamsi, R.~Alizadehsani, A.~Khosravi, S.~Nahavandi, Z.~A.
  Sani \emph{et~al.}, ``Objective evaluation of deep uncertainty predictions
  for {COVID}-19 detection,'' 2020.

\bibitem{Kourou2015}
K.~Kourou, T.~P. Exarchos, K.~P. Exarchos, M.~V. Karamouzis, and D.~I.
  Fotiadis, ``Machine learning applications in cancer prognosis and
  prediction,'' \emph{Computational and Structural Biotechnology Journal},
  vol.~13, pp. 8--17, 2015.

\bibitem{Choy2018}
G.~Choy, O.~Khalilzadeh, M.~Michalski, S.~Do, A.~E. Samir, O.~S. Pianykh
  \emph{et~al.}, ``Current applications and future impact of machine learning
  in radiology,'' \emph{Radiology}, vol. 288, no.~2, pp. 318--328, 2018, pMID:
  29944078.

\bibitem{Deo2015}
R.~C. Deo, ``Machine learning in medicine,'' \emph{Circulation}, vol. 132,
  no.~20, pp. 1920--1930, 2015.

\bibitem{Lee2017}
J.-G. Lee, S.~Jun, Y.-W. Cho, H.~Lee, G.~B. Kim, J.~B. Seo \emph{et~al.},
  ``Deep learning in medical imaging: General overview,'' \emph{Korean J
  Radiol}, vol.~18, no.~4, pp. 570--584, 2017.

\bibitem{Kallianos2019}
K.~Kallianos, J.~Mongan, S.~Antani, T.~Henry, A.~Taylor, J.~Abuya
  \emph{et~al.}, ``How far have we come? artificial intelligence for chest
  radiograph interpretation,'' \emph{Clinical Radiology}, vol.~74, no.~5, pp.
  338--345, 2019.

\bibitem{Monshi2020}
M.~M.~A. Monshi, J.~Poon, and V.~Chung, ``Deep learning in generating radiology
  reports: A survey,'' \emph{Artificial Intelligence in Medicine}, vol. 106, p.
  101878, 2020.

\bibitem{Fellous2019}
J.-M. Fellous, G.~Sapiro, A.~Rossi, H.~Mayberg, and M.~Ferrante, ``Explainable
  artificial intelligence for neuroscience: Behavioral neurostimulation,''
  \emph{Frontiers in Neuroscience}, vol.~13, p. 1346, 2019.

\bibitem{He2019}
J.~He, S.~L. Baxter, J.~Xu, J.~Xu, X.~Zhou, and K.~Zhang, ``The practical
  implementation of artificial intelligence technologies in medicine,''
  \emph{Nature Medicine}, vol.~25, no.~1, pp. 30--36, 2019.

\bibitem{Holzinger2017}
A.~Holzinger, C.~Biemann, C.~S. Pattichis, and D.~B. Kell, ``What do we need to
  build explainable {AI} systems for the medical domain?'' 2017.

\bibitem{Holzinger2019}
A.~Holzinger, G.~Langs, H.~Denk, K.~Zatloukal, and H.~Müller, ``Causability
  and explainability of artificial intelligence in medicine,'' \emph{WIREs Data
  Mining and Knowledge Discovery}, vol.~9, no.~4, p. e1312, 2019.

\bibitem{Litjens2017}
G.~Litjens, T.~Kooi, B.~E. Bejnordi, A.~A.~A. Setio, F.~Ciompi, M.~Ghafoorian
  \emph{et~al.}, ``A survey on deep learning in medical image analysis,''
  \emph{Medical Image Analysis}, vol.~42, pp. 60--88, 2017.

\bibitem{Yu2018}
K.-H. Yu, A.~L. Beam, and I.~S. Kohane, ``Artificial intelligence in
  healthcare,'' \emph{Nature Biomedical Engineering}, vol.~2, no.~10, pp.
  719--731, 2018.

\bibitem{Altaf2019}
F.~Altaf, S.~M.~S. Islam, N.~Akhtar, and N.~K. Janjua, ``Going deep in medical
  image analysis: Concepts, methods, challenges, and future directions,''
  \emph{IEEE Access}, vol.~7, pp. 99\,540--99\,572, 2019.

\bibitem{Mongan2020}
\BIBentryALTinterwordspacing
J.~Mongan, L.~Moy, and C.~E. Kahn, ``Checklist for artificial intelligence in
  medical imaging ({CLAIM}): A guide for authors and reviewers,''
  \emph{Radiology: Artificial Intelligence}, vol.~2, no.~2, p. e200029, 2020,
  pMID: 33937821. [Online]. Available:
  \url{https://doi.org/10.1148/ryai.2020200029}
\BIBentrySTDinterwordspacing

\bibitem{Lambin2017}
\BIBentryALTinterwordspacing
P.~Lambin, R.~T.~H. Leijenaar, T.~M. Deist, J.~Peerlings, E.~E.~C. de~Jong,
  J.~van Timmeren \emph{et~al.}, ``Radiomics: the bridge between medical
  imaging and personalized medicine,'' \emph{Nature Reviews Clinical Oncology},
  vol.~14, no.~12, pp. 749--762, 2017. [Online]. Available:
  \url{https://doi.org/10.1038/nrclinonc.2017.141}
\BIBentrySTDinterwordspacing

\bibitem{Collins2015}
G.~S. Collins, J.~B. Reitsma, D.~G. Altman, and K.~G.~M. Moons, ``Transparent
  reporting of a multivariable prediction model for individual prognosis or
  diagnosis ({TRIPOD}): the {TRIPOD} statement.'' \emph{BMJ (Clinical research
  ed.)}, vol. 350, no. jan07 4, p. g7594, 2015.

\bibitem{Wolff2019}
R.~F. Wolff, K.~G.~M. Moons, R.~D. Riley, P.~F. Whiting, M.~Westwood, G.~S.
  Collins \emph{et~al.}, ``{PROBAST}: A tool to assess the risk of bias and
  applicability of prediction model studies,'' \emph{Ann Intern Med}, vol. 170,
  no.~1, pp. 51--58, 2019.

\bibitem{Castro2020}
D.~C. Castro, I.~Walker, and B.~Glocker, ``Causality matters in medical
  imaging,'' \emph{Nature Communications}, vol.~11, no.~1, p. 3673, 2020.

\bibitem{Almasni2020}
M.~A. Al-masni, D.-H. Kim, and T.-S. Kim, ``Multiple skin lesions diagnostics
  via integrated deep convolutional networks for segmentation and
  classification,'' \emph{Computer Methods and Programs in Biomedicine}, vol.
  190, p. 105351, 2020.

\bibitem{Subbaswamy2019}
A.~Subbaswamy and S.~Saria, ``From development to deployment: dataset shift,
  causality, and shift-stable models in health {AI},'' \emph{Biostatistics},
  vol.~21, no.~2, pp. 345--352, 2019.

\bibitem{Badgeley2019}
M.~A. Badgeley, J.~R. Zech, L.~Oakden-Rayner, B.~S. Glicksberg, M.~Liu, W.~Gale
  \emph{et~al.}, ``Deep learning predicts hip fracture using confounding
  patient and healthcare variables,'' \emph{npj Digital Medicine}, vol.~2,
  no.~1, p.~31, 2019.

\bibitem{Oala2020}
\BIBentryALTinterwordspacing
L.~Oala, J.~Fehr, L.~Gilli, P.~Balachandran, A.~W. Leite, S.~Calderon-Ramirez
  \emph{et~al.}, ``{ML4H} auditing: From paper to practice,'' in
  \emph{Proceedings of the Machine Learning for Health NeurIPS Workshop}, ser.
  Proceedings of Machine Learning Research, E.~Alsentzer, M.~B.~A. McDermott,
  F.~Falck, S.~K. Sarkar, S.~Roy, and S.~L. Hyland, Eds., vol. 136.\hskip 1em
  plus 0.5em minus 0.4em\relax PMLR, 2020, pp. 280--317. [Online]. Available:
  \url{http://proceedings.mlr.press/v136/oala20a.html}
\BIBentrySTDinterwordspacing

\bibitem{fg-ai4h}
\BIBentryALTinterwordspacing
{FG-AI4H}, ``The {ITU/WHO} focus group on artificial intelligence for health,''
  2021. [Online]. Available:
  \url{https://www.itu.int/en/ITU-T/focusgroups/ai4h/Pages/default.aspx}
\BIBentrySTDinterwordspacing

\bibitem{Leli2021}
C.~Leli, L.~{Di Matteo}, F.~Gotta, E.~Cornaglia, D.~Vay, I.~Megna
  \emph{et~al.}, ``Performance of a {SARS-CoV-2} antigen rapid immunoassay in
  patients admitted to the emergency department,'' \emph{International Journal
  of Infectious Diseases}, vol. 110, pp. 135--140, 2021.

\bibitem{Shorten2019}
\BIBentryALTinterwordspacing
C.~Shorten and T.~M. Khoshgoftaar, ``A survey on image data augmentation for
  deep learning,'' \emph{Journal of Big Data}, vol.~6, no.~1, p.~60, 2019.
  [Online]. Available: \url{https://doi.org/10.1186/s40537-019-0197-0}
\BIBentrySTDinterwordspacing

\bibitem{Stephen2019}
\BIBentryALTinterwordspacing
O.~Stephen, M.~Sain, U.~J. Maduh, and D.-U. Jeong, ``An efficient deep learning
  approach to pneumonia classification in healthcare,'' \emph{Journal of
  Healthcare Engineering}, vol. 2019, p. 4180949, 2019. [Online]. Available:
  \url{https://doi.org/10.1155/2019/4180949}
\BIBentrySTDinterwordspacing

\bibitem{Goodfellow2014}
\BIBentryALTinterwordspacing
I.~J. Goodfellow, J.~Pouget-Abadie, M.~Mirza, B.~Xu, D.~Warde-Farley, S.~Ozair
  \emph{et~al.}, ``Generative adversarial nets,'' in \emph{NIPS}, 2014, pp.
  2672--2680. [Online]. Available:
  \url{http://papers.nips.cc/paper/5423-generative-adversarial-nets}
\BIBentrySTDinterwordspacing

\bibitem{Weiss2016}
\BIBentryALTinterwordspacing
K.~Weiss, T.~M. Khoshgoftaar, and D.~Wang, ``A survey of transfer learning,''
  \emph{Journal of Big Data}, vol.~3, no.~1, p.~9, 2016. [Online]. Available:
  \url{https://doi.org/10.1186/s40537-016-0043-6}
\BIBentrySTDinterwordspacing

\bibitem{Tan2018}
C.~Tan, F.~Sun, T.~Kong, W.~Zhang, C.~Yang, and C.~Liu, ``A survey on deep
  transfer learning,'' in \emph{Artificial Neural Networks and Machine Learning
  -- ICANN 2018}, V.~K{\r{u}}rkov{\'a}, Y.~Manolopoulos, B.~Hammer, L.~Iliadis,
  and I.~Maglogiannis, Eds.\hskip 1em plus 0.5em minus 0.4em\relax Cham:
  Springer International Publishing, 2018, pp. 270--279.

\bibitem{Abbas2020}
A.~Abbas, M.~M. Abdelsamea, and M.~M. Gaber, ``{DeTrac}: Transfer learning of
  class decomposed medical images in convolutional neural networks,''
  \emph{IEEE Access}, vol.~8, pp. 74\,901--74\,913, 2020.

\bibitem{Karimi2021}
\BIBentryALTinterwordspacing
D.~Karimi, S.~K. Warfield, and A.~Gholipour, ``Transfer learning in medical
  image segmentation: New insights from analysis of the dynamics of model
  parameters and learned representations,'' \emph{Artificial Intelligence in
  Medicine}, vol. 116, p. 102078, 2021. [Online]. Available:
  \url{https://www.sciencedirect.com/science/article/pii/S0933365721000713}
\BIBentrySTDinterwordspacing

\bibitem{Raghu2019}
\BIBentryALTinterwordspacing
M.~Raghu, C.~Zhang, J.~Kleinberg, and S.~Bengio, ``Transfusion: Understanding
  transfer learning for medical imaging,'' in \emph{Advances in Neural
  Information Processing Systems}, H.~Wallach, H.~Larochelle, A.~Beygelzimer,
  F.~d\textquotesingle Alch\'{e}-Buc, E.~Fox, and R.~Garnett, Eds.,
  vol.~32.\hskip 1em plus 0.5em minus 0.4em\relax Curran Associates, Inc.,
  2019. [Online]. Available:
  \url{https://proceedings.neurips.cc/paper/2019/file/eb1e78328c46506b46a4ac4a1e378b91-Paper.pdf}
\BIBentrySTDinterwordspacing

\bibitem{Cheplygina2019}
\BIBentryALTinterwordspacing
V.~Cheplygina, ``Cats or {CAT} scans: Transfer learning from natural or medical
  image source data sets?'' \emph{Current Opinion in Biomedical Engineering},
  vol.~9, pp. 21--27, 2019. [Online]. Available:
  \url{https://www.sciencedirect.com/science/article/pii/S2468451118300527}
\BIBentrySTDinterwordspacing

\end{thebibliography}

\end{document}